# The First Results of Distributed Peer Review at ESO Show Promising Outcomes

Tereza Jerabkova [1]
Ferdinando Patat [1]
Francesca Primas [1]
Dario Dorigo [1]
Fabio Sogni [1]
Lucas Astolfi [1]
Thomas Bierwirth [1]
Michael Prümm [1]

[1] ESO

**The European Southern Observatory (ESO) implemented a new paradigm called Distributed Peer Review (DPR) as part of its proposal evaluation process in Period 110. Under DPR, Principal Investigators who submit proposals agree to review a certain number of proposals submitted by their peers and accept that their own proposal(s) are reviewed by their peers who have also submitted proposals in the same cycle. This article presents a brief overview of the DPR process at ESO, and its outcomes based on data from periods 110 and 111.**

## DPR introduction and ESO DPR P110 and P111 overview

The Distributed Peer Review (DPR) paradigm should be seen not only as an innovative concept, but above all as a natural consequence of the increased number of proposals requiring review. Different opinions on whether the expert panel review model needed revision have been put forward more than once since its inception a couple of centuries back. However, the recently increased numbers of applications for observing time made this issue more pressing and challenging for all large astronomical facilities and their time allocation processes. To keep logistical aspects manageable and, at the same time, to ensure a high-quality proposal evaluation, peer review by standard expert panels has become a less and less viable option, calling for suitable and sustainable alternatives.

The DPR concept was first introduced and formalised by Merrifield & Saari (2009). The general idea underlying it is that submitted proposals are not reviewed by pre-selected expert panels but rather by fellow PIs who have also applied for observing time, hence actually staying true to the very concept of peer review, in which the applicants and the referees are at the same level. Gemini Observatory pioneered the adoption of DPR for telescope time allocation in their Fast Turnaround channel in 2015 (Andersen, 2020). DPR was first considered at ESO following a report by the Time Allocation Working Group (Patat, 2018), which turned into a successful DPR experiment that was conducted on a voluntary basis alongside the panel review (Patat et al., 2019). DPR was deployed for ALMA time allocation in 2021 for Cycle 8 (Meyer et al., 2022), followed by ESO in 2022 for Period 110. In this article we present a statistical analysis of the first two semesters.

In P110 and P111 proposals requesting more than 16 hours of observing time and special programmes (for example, Targets of Opportunity, Calibration Proposals, and joint XMM-Newton proposals) were evaluated by expert panels and approved by the Observing Programmes Committee (OPC). The 16-hour limit was chosen to produce a balanced distribution between DPR and panels, effectively reducing the load on the panels by a factor of two, while leaving about 80% of the time under their control. The threshold value was defined in the Call for Proposals for the corresponding period, and it might be adjusted in future. The DPR channel evaluated 435 proposals distributed to 379 reviewers in P110 and 417 proposals with 362 reviewers in P111.

The DPR scheme was set up so that PIs (or delegated PIs) submitting proposals that qualified for DPR had to agree — at time of submission — to evaluate 10 proposals per submitted (DPR) proposal. They also had the option of selecting one of the proposal co-Is as the delegated DPR reviewer. Failing to do this by the set deadline would result in the automatic rejection of their proposal/s. This guaranteed that each proposal received 10 independent grades (from 1 to 5) and comments.

## Expertise evaluation and proposal assignment

An important aspect of any peer review process is the set of criteria that are used to assign proposals to the reviewers. The ESO DPR in P110 and P111 aimed for expert peer review in which proposals were assigned to reviewers with expertise as close as possible to the science case of the proposals. Thus, the expertise level for each proposal-reviewer pair needed to be evaluated and the assignment algorithm was then responsible for an optimal proposal distribution that maximised the match overall.

### Reviewer-proposal expertise score

The Phase 1 tool (P1)[1] introduced by Primas et al. (2019) has revolutionised the way proposals are submitted to ESO. In order to submit a proposal, users must have a User Portal account, where they must provide information about their career stage, affiliation, and scientific expertise through keywords. These keywords are grouped into classes such as Cosmology, Galaxies and Galactic Nuclei, Interstellar Medium, Star Formation and Planetary Systems, and Stellar Evolution. Each proposal submitted through the P1 tool must include a subset of these keywords that describe the proposal's science scope (up to five and a minimum of three).

In each case, the keywords are specified by the applicant in decreasing order of relevance. This information is then used to compose a knowledge vector for each proposal and each reviewer. For each reviewer-proposal pair, it is then possible to compute the scalar product of their respective knowledge vectors, resulting in what we will refer to as the match score. In other words, the match score is a figure of merit describing how parallel the two vectors are. The match scores range from 0 (no match), through 1 (match within a science category), to 2 (perfect full match). This information provides essential input for the assignment process and is used to ensure that proposals are assigned in a controlled way to the most qualified reviewers.

### Proposal-reviewer matching

The proposal assignment is a thorough and methodical process that takes into account various factors to ensure the best possible match between proposals and reviewers. One of the key factors considered is the match score of each proposal-reviewer pair introduced above. To ensure fairness and consistency, ESO has specific rules for assigning proposals. For example, each reviewer is assigned a set number of proposals based on the number of submissions, and each proposal is assigned to a set number of reviewers. In addition, possible conflicts between reviewers, such as those arising from team or institute membership, are also considered, to avoid any potential bias. To manage the assignment process, ESO uses a java web-based tool called proposal_distributor, which employs an algorithm developed in-house to sort proposals based on their assignability. Proposals with the smallest pool of possible reviewers are given priority above the proposals with a larger number of available experts. Reviewers are then assigned to proposals based on their expertise scores and the prior sorting. This process is repeated until all the assignment rules and constraints are met. The tool is designed to be efficient and transparent and ensures the best match between proposals and reviewers.

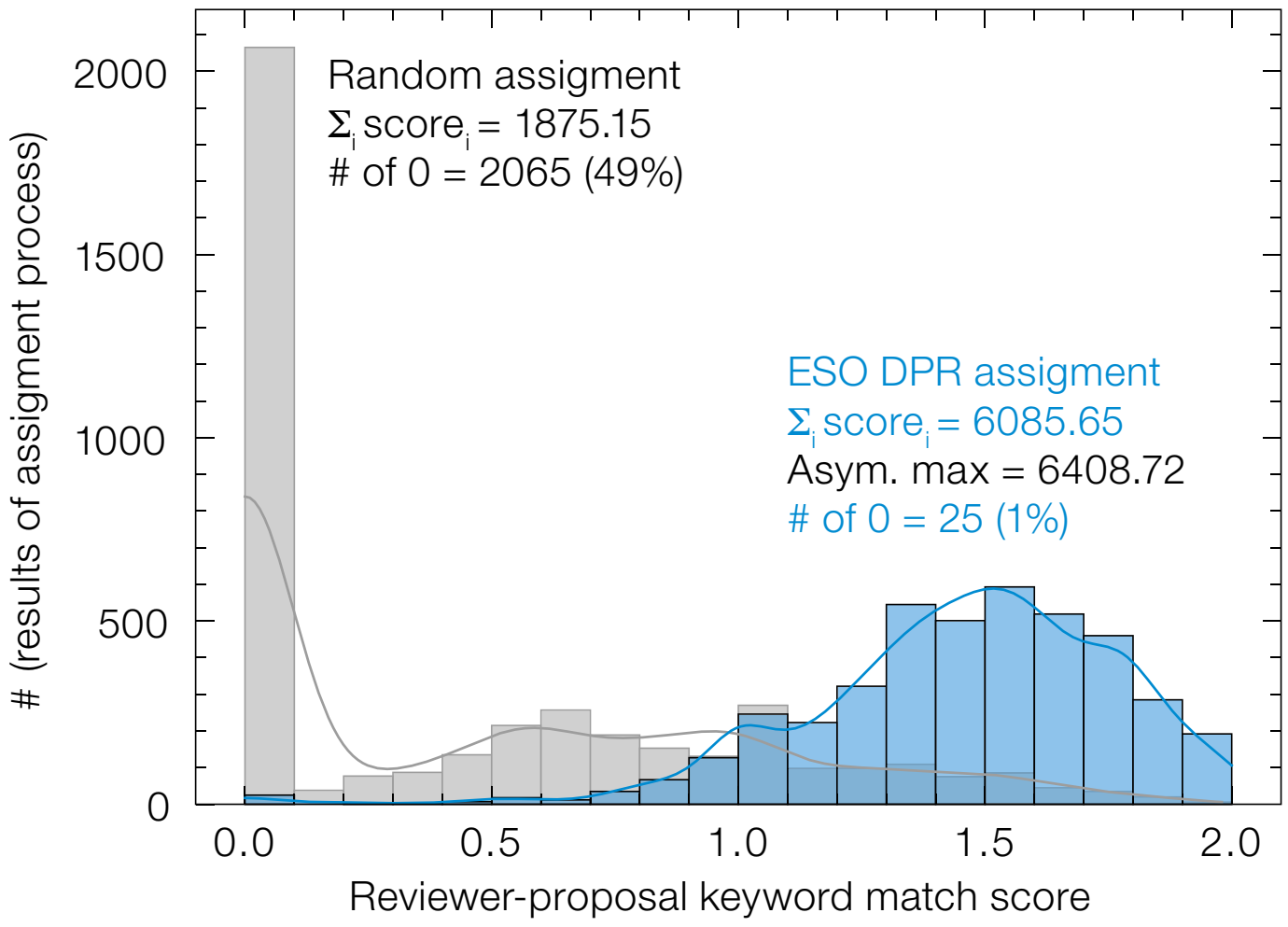

**Figure 1.** Example proposal assignment outcome in P111 (P110 provides quantitatively the same figure). The horizontal axis shows the match score, and the vertical axis shows the number of proposals assigned in each score bin. The blue histogram shows the assignment produced by the proposal_distributor. In grey, we show a random assignment that is representative of the overall score distribution.

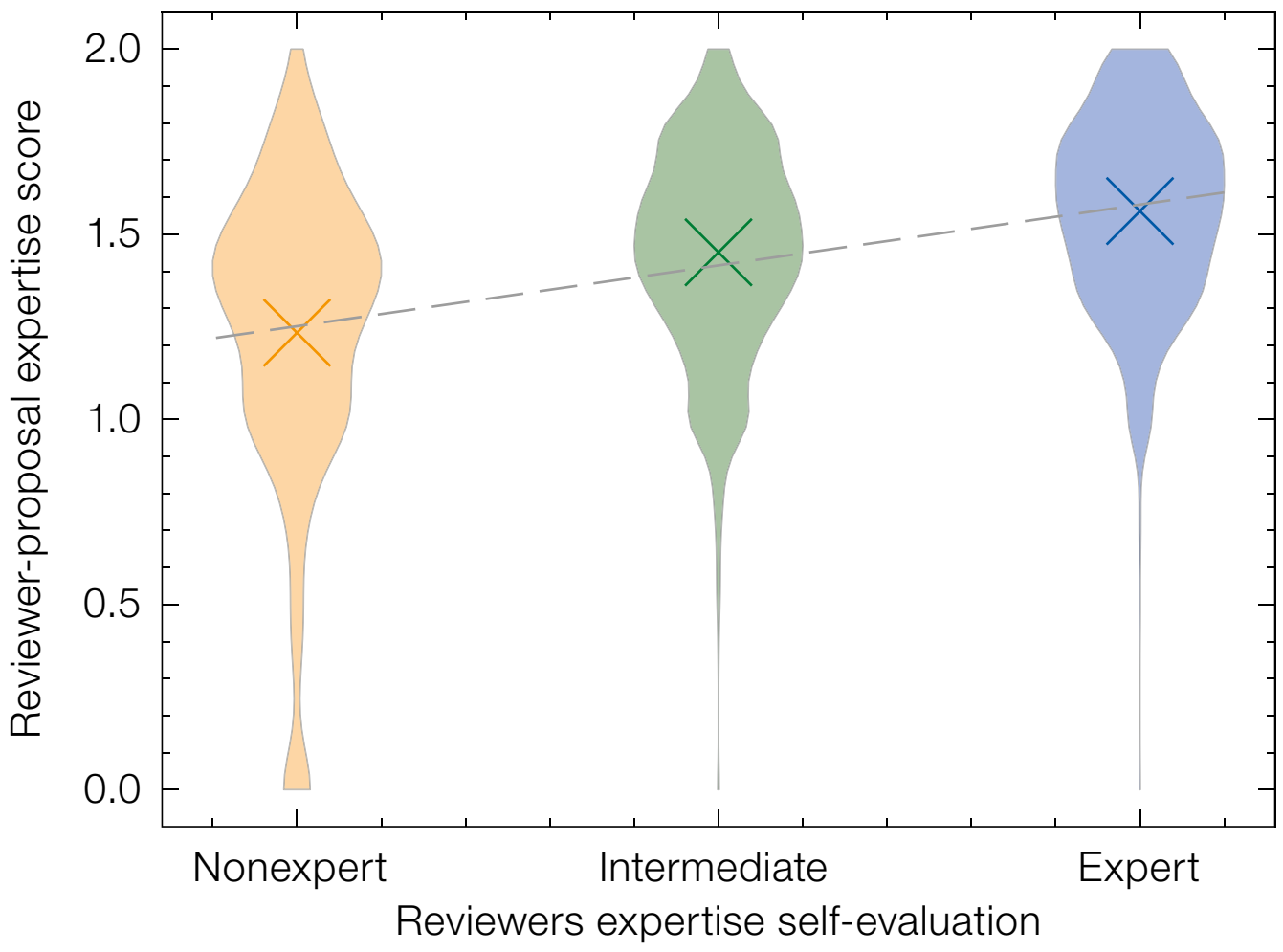

**Figure 2.** The reviewer-proposal match score calculated based on specified keywords compared to reviewers' expertise self-evaluation.

Figure 1 shows the proposal assignment outcome in blue. The shape of the distribution peaks towards high scores, showing the effectiveness of the assignment algorithm. This can be compared to the distribution of all scores, that is, the raw input for the assignment algorithm. For this we have plotted in grey a random assignment for the same number of proposals/referees. One straightforward way of quantifying the quality of the overall assignment outcome is to sum all their match scores. A higher number means a better assignment in terms of expertise, while a null score means there are no common keywords for the reviewer-proposal pair. The final score value for the proposal_distributor is more than three times larger than in the case of the random assignment. Another comparison to gauge the performance of the matching algorithm can be made with the best distribution one can have with the available set of reviewers and proposals. To this end, for a given distribution of scores, one can construct a hypothetical asymptotic assignment which maximises the final score. This ideal proposal distribution takes the ten best assignments for each proposal, excluding conflicts, but ignoring the boundary constraint that each reviewer can have only 10 proposals to assess. This therefore represents an asymptotic, practically unachievable, upper limit for the match. The asymptotic score is a factor of 3.4 larger than the random assignment outcome and only a factor of 1.05 larger when compared to the proposal_distributor.

## P110 und P111 summary

The DPR process in P110 and P111 ran smoothly, no technical issues were reported, and all reviewers delivered their evaluations on time. The final grade distribution was carefully analysed, showing that the DPR reviewers behaved statistically as panel members prior to the discussion phase. In addition, we did not detect any statistically significant systematic effects related to seniority or career stage. To produce the final proposal ranking list the DPR grades were normalised to have the same mean value and standard deviation as the grades awarded by the panels, adopting the same procedure in place for aligning the outcomes of individual panels.

## Match scores and reviewers' self-evaluation

As discussed in the previous section, the match scores are a crucial ingredient of the DPR process. In the current system, we rely fully on the users and their input keywords, as specified in the ESO User Portal profiles and in the submitted pro-posals. To monitor the performance of the algorithm, the DPR proposal evaluation interface requires each reviewer to express their self-evaluated expertise level (non-expert, intermediate, expert) for each proposal assigned to them. Figure 2 shows a comparison between the reviewer-proposal match score and the self-evaluated expertise level. For each self-evaluated expertise value, the match score distribution is plotted, with its mean value denoted by a cross. There is an overall trend, indicated by a dashed line fitted to the mean values, showing a positive correlation between the two indicators. Typically, low match scores are consistently evaluated as 'non-expert' by the users. However, several users categorised themselves as 'non-expert' while their match score was as high as 2, thus indicating a perfect keyword match (the same keywords specified in exactly the same order). These cases can have several possible explanations, such as incorrectly assigned keywords (either in the User Profile or in the proposal), keywords that were too broad, or subjective bias. In the data collected in P110, we noticed that early-career scientists were less likely to claim to be an expert, even in their own field. On the other hand, senior scientists, such as professors and staff astronomers, were more likely to claim to be an expert even in a field that is far from their declared expertise.

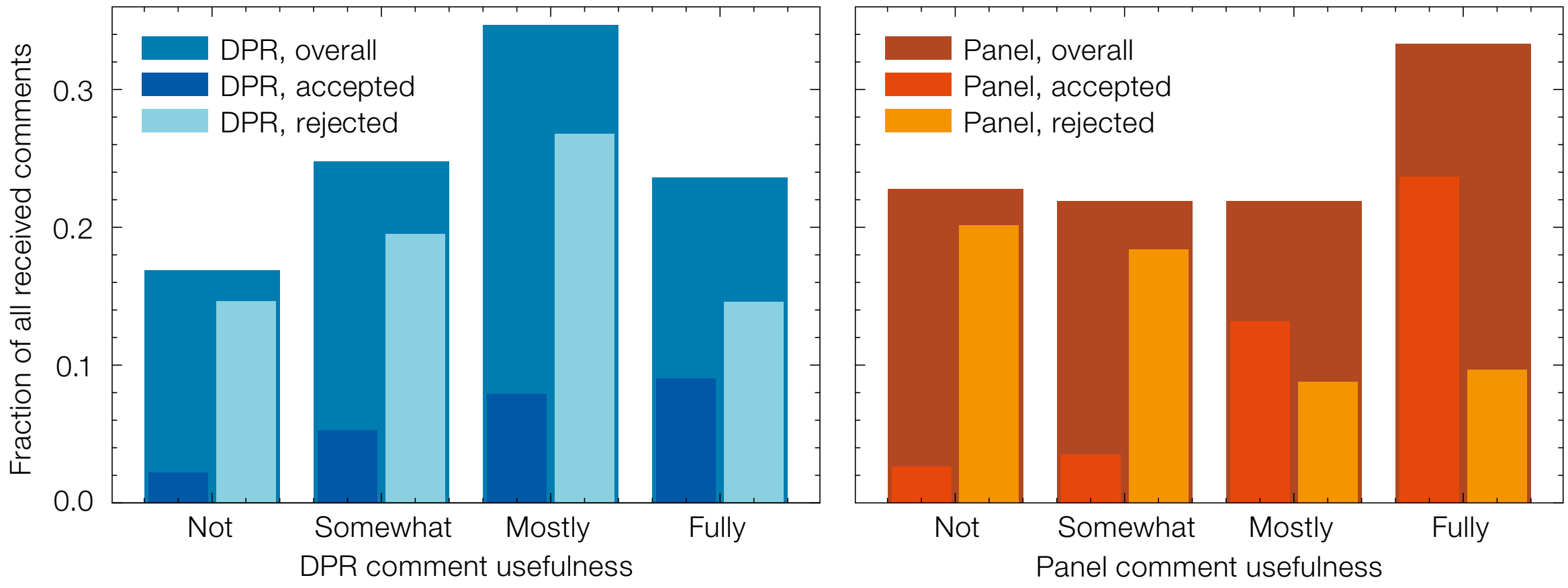

**Figure 3.** The comment usefulness as declared by PIs/dPIs in Period 110 for DPR (left) and panels (right).

## Feedback on the reviewers' proposal comments

In addition to grading proposals, both in DPR and panels, the reviewers are responsible for providing useful feedback to the PIs. This feedback should clearly point out the strengths and weaknesses of the proposal and provide indications for improving it when applicable. For the first time in P110, we invited ESO PIs/dPIs to evaluate the usefulness of the received comments, on a voluntary basis. The PIs' response rate was the same for both the expert panels and DPR: around 30% provided their evaluation of the feedback they received. This fraction is not very satisfactory, and measures to increase it in future semesters are under discussion. However, the data already allow some basic analysis.

Figure 3 shows the distribution of the collected responses for DPR (left) and panels (right). These first results indicate a higher level of satisfaction for DPR. This becomes even more evident when looking only at the rejected proposals, which are arguably the most relevant cases for this specific aspect. While in the case of the panels most of the comments are judged as 'not' or 'somewhat' useful, an opposite trend is seen in the case of DPR. When considering this result, one must emphasise that the panels produce one single joint comment for each proposal, emerging from the discussion at the meeting, while in the case of DPR, each reviewer writes an independent evaluation, which is passed directly and unedited to the PI. Thus, in the case of DPR the feedback on the proposals captures more information, which is the likely explanation for the observed trend.

## Summary and future prospects

DPR was deployed smoothly at ESO in periods P110 and P111, and the outcome is statistically comparable to that of the expert panels. The user feedback on the usefulness of the comments received by the PIs clearly favours DPR. In this respect, it is worth noticing that the decreasing level of user satisfaction with the feedback provided by the panels was one of the most quantitative drivers for considering DPR as a possible alternative to the classical paradigm. It is reassuring to see that DPR has improved this situation. At the same time, the data seem to indicate that, following the introduction of DPR, the level of satisfaction with the panels' feedback has increased, a beneficial result of the significant decrease in the number of proposals reviewed by each panel member.

As recommended by ESO's Scientific and Technical Committee, we will keep the same setup for P112 and P113, to collect more information under the same homogeneous conditions. Nevertheless, we are planning to further monitor and improve the process and to keep presenting our findings in order to remain as transparent as possible. More complex assignment algorithms are being tested (for example, Faez, Dickerson & Fuge, 2017; Stelmakh, Shah & Singh, 2018) and we are working on improving reviewer and proposal profiling via machine learning approaches (Patat et al., 2019; Kerzendorf et al., 2020). We are also looking into ways of increasing the rate of feedback return from the PIs, in order to have a larger statistical basis for future analyses.

As part of the development process, we organised the conference Peer-Review Under Review, held at ESO's Headquarters in Garching, Germany, on 6–10 February 2023, the outcome of which will be presented in a future Messenger article. The idea was to bring together not only the astronomers, but also representatives of the wider scientific community, in order to start a discussion about peer-review processes in general, and to identify ways to cope with a continuously growing scientific community and the appearance of intriguing new technologies.

Back in 1973, in his science fiction novel *Rendezvous with Rama*, Arthur C. Clarke wrote prophetically about an expert panel evaluating a proposal for a space mission: *"Even by the twenty-second century, no way had yet been discovered of keeping elderly and conservative scientists from occupying crucial administrative positions. Indeed, it was doubted if the problem ever would be solved."* The promising outcome of the deployment of DPR at ALMA and ESO indicates that this may indeed be one viable and valid alternative to the standard expert panel review process. For once, we might be ahead of science fiction.

## Acknowledgements

The authors are grateful to ESO's Director General and Director for Science and to Markus Kissler-Patig for supporting this enterprise, which represents the most significant change in the proposal evaluation process at ESO since it was established. The support and the positive attitude shown by the Scientific Technical Committee, the Users Committee and the Observing Programmes Committee are also acknowledged. We would like to acknowledge the valuable assistance of OpenAI's language model, ChatGPT.

## Links

[1] The ESO P1 tool: https://www.eso.org/sci/observing/phase1/p1intro.html

ENEL

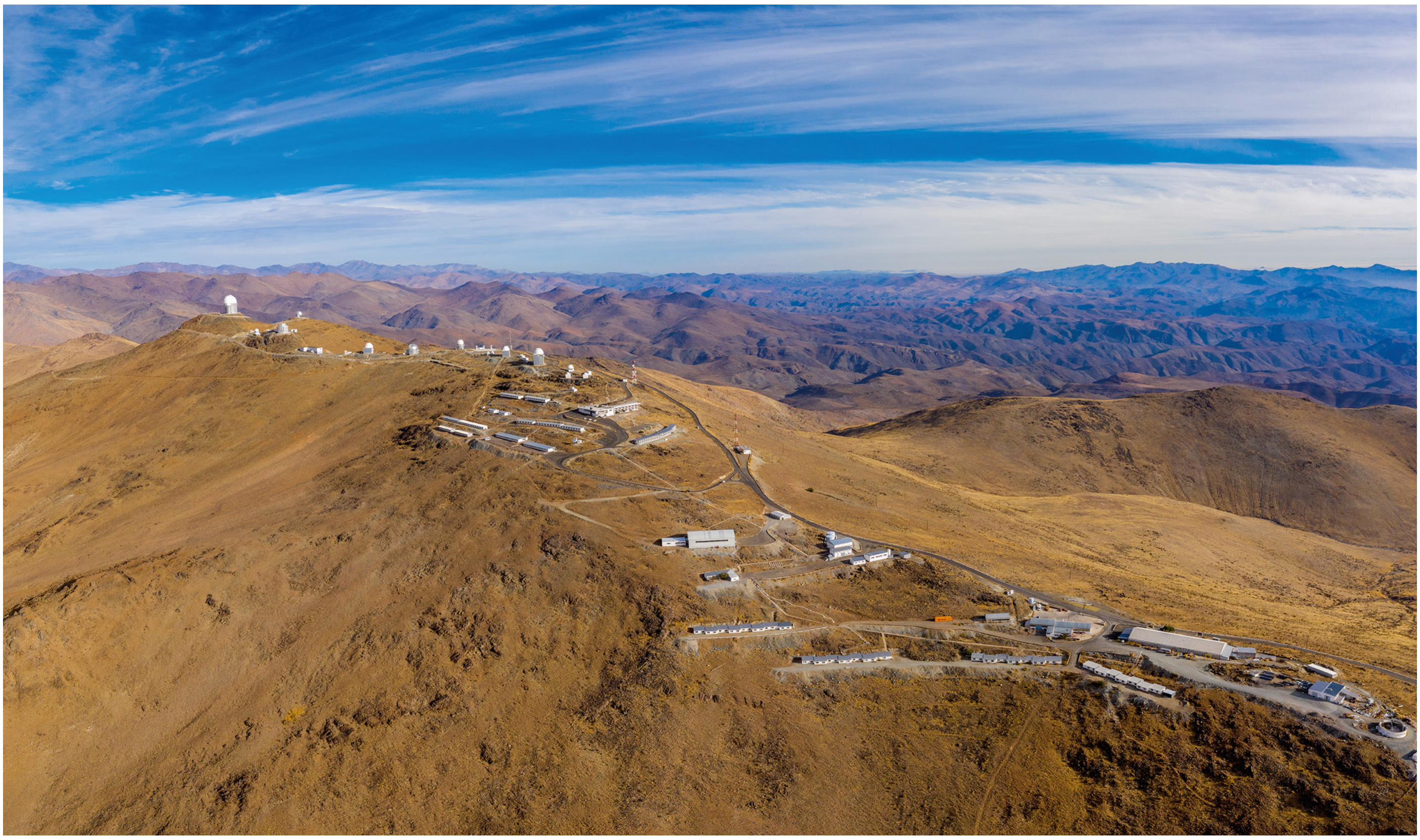

Located on the outskirts of the Chilean Atacama Desert, 600 km north of Santiago and at an altitude of 2400 metres, this seemingly tiny village in the middle of a desert is in fact ESO's first observatory, La Silla Observatory.